\documentclass[sigconf]{acmart}
\usepackage{float}
\usepackage{booktabs}
\usepackage{comment}
\usepackage{url}
\usepackage{xurl}
\usepackage{algorithmic}
\usepackage{graphicx}
\usepackage{textcomp}
\def\BibTeX{{\rm B\kern-.05em{\sc i\kern-.025em b}\kern-.08em
    T\kern-.1667em\lower.7ex\hbox{E}\kern-.125emX}}
\usepackage[table]{xcolor} 
\definecolor{mygray}{rgb}{0.83, 0.83, 0.83}
\newcommand{\cut}[1]{}
\usepackage[inline]{enumitem}
\usepackage{multicol}
\usepackage{multirow}
\usepackage{makecell}
\setlist[enumerate,1]{label=\alph*)}
\usepackage{siunitx} 
\usepackage{makecell}
\usepackage{threeparttable}
\usepackage{tabularx}
\usepackage[normalem]{ulem}
\usepackage{cleveref}
\usepackage{csquotes}
\usepackage{pdflscape} 
\usepackage{graphicx}
\usepackage[font=small]{caption}
\newcommand{\quotes}[1]{``#1''} 
 
\newcommand{\autorefappendix}[1]{\hyperref[#1]{Appendix~\ref*{#1}}}  
\crefname{appendix}{appendix}{appendices}
\Crefname{appendix}{Appendix}{Appendices}

\usepackage{tcolorbox}
\setcopyright{none}
\renewcommand\footnotetextcopyrightpermission[1]{} 

\acmConference[]{}{}{}
\acmYear{}

\begin{document}

\title[]{Collaborative Hand Drawing to Conceptualize Usable Privacy and Security Research Problems: Students' Perspectives (Extended Version)}

\author{Eman Alashwali}
\authornote{Corresponding author and principal investigator.}
\affiliation{%
  \institution{King Abdulaziz University}
  \city{Jeddah}
  \country{Saudi Arabia}
}
\email{ealashwali@kau.edu.sa}

\author{Joud Alraimi}
\authornote{Authors contributed equally to this work.}
\affiliation{%
  \institution{King Abdulaziz University}
  \city{Jeddah}
  \country{Saudi Arabia}
}
\email{jalraimi@stu.kau.edu.sa}

\author{Maryam Almaghrabi}
\authornotemark[2]
\affiliation{%
  \institution{King Abdulaziz University}
  \city{Jeddah}
  \country{Saudi Arabia}
}
\email{malmaghrabi0059@stu.kau.edu.sa}
\renewcommand{\shortauthors}{Alashwali et al.}

\begin{abstract}
In this paper, we explore students' perspectives on incorporating hand drawing to conceptualize usable privacy and security research problems. We surveyed 64 female students who participated in an in-class collaborative hand drawing activity to conceptualize usable privacy and security research problems as part of a group-based student research project on usable privacy and security in an information and computer security course for undergraduate students. Our survey responses suggest that students perceived the drawing exercise positively. That is, they found it helpful in several aspects, enjoyable, and they recommend it to other students taking the course. The students also said that they would like to have such an exercise in other courses, such as Principles of Human-Computer Interaction and Fundamentals of Data Networks. Only three students mentioned downsides of the exercise, namely doubts about their drawing skills. Our results provide useful insights for educators seeking to enhance the student learning experience. This paper is an extended version of our previously published short paper~\cite{alashwali26}.
\end{abstract}


\maketitle
\pagestyle{plain}
\begin{tcolorbox}
This document is the authors' extended preprint for a short paper in the proceedings of Koli Calling'26, ACM (DOI: 10.1145/3856208.3856277).
\end{tcolorbox}
\section{Introduction}
\begin{figure*}
    \centering
    \includegraphics[width=0.8\linewidth]{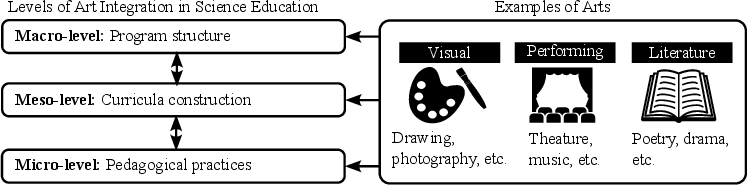}
    \caption{A layered overview of the levels at which art can be integrated into science education as described in Braund and Reiss~\cite{braund19}. The meso-level intersects with the macro and micro levels, indicated by the double-headed arrows. The arts included in the right side are examples and are not inclusive of all types of arts.}
    \label{fig:steam}
\end{figure*}

In recent years, there has been growing interest in developing the relationship between art and science, suggesting that art complements science and science education~\cite{braund19}. This has been evident in initiatives advocating for STEAM instead of STEM~\cite{braund19}. Namely, adding \enquote{Arts} to \enquote{Science, Technology, Engineering and Mathematics}~\cite{yakman08,braund19}. As depicted in~\Cref{fig:steam}, Braud and Reiss suggested that art can be incorporated in science education at three levels: \begin{enumerate*} [label=(\arabic*)] \item macro-level (program structure and course options), \item meso-level (curriculum construction through Science, Technology, Society and Environment (STSE) contexts~\cite{pedretti11}), and \item micro-level (science pedagogical practices that incorporate arts)~\cite{braund19}.\end{enumerate*} Arts can be integrated into science education in various forms, such as visual arts (e.g., drawing), literature (e.g., poetry), and performing arts (e.g., music)~\cite{braund19,segarra18}. Accordingly, our incorporation of hand drawing exercises into a course research project is considered at the micro level.  

In the learning context, Quillin and Thomas provided the following definition for drawing~\cite{quillin15}: 
\textit{a learner-generated external visual representation depicting any type of content, whether structure, relationship, or process, created in static two dimensions in any medium.}

Plenty of studies on drawing and learning suggest that drawing can foster various pedagogical goals, such as motivation, reasoning, invention, action, communication, learning to represent, improving engagement, and enjoyment of learning~\cite{ainsworth11,quillin15,adams17}. 

This paper is mainly concerned with hand drawing as a means of assisting the development of ideas. The main research question that this study aims to answer is: 
\begin{quote}
\textit{RQ: Whether and how does incorporating a hand drawing exercise help students in conceptualizing a usable privacy and security research problem?}
\end{quote}

To this end, we surveyed female students who participated in a collaborative hand drawing exercise to conceptualize a usable privacy and security research problem as part of a group-based student research project in the Information and Computer Security course taught by the first author to Information Systems Department students at the Faculty of Computing and Information Technology, King Abdulaziz University, Saudi Arabia.

Our results support incorporating hand drawing exercises to conceptualize usable privacy and security research problems. The drawing exercise was perceived positively by students in all examined aspects. It was helpful, enjoyable, and students recommended the drawing exercise to other students in the same course. In terms of helpfulness, we asked about several aspects. Students found the exercise helpful in all examined aspects, though in various degrees. The exercise was most helpful in developing positive team relationship, motivating them to work in the project, clarifying the conception of the project idea, developing the research question, and finally fostering a sense of project ownership. Students also would like to have hand drawing exercises in other courses, and they named courses where they perceive drawing can be helpful.   

Our experience and findings represent a case study that can help improve educational methods in Computer Science (CS), particularly cybersecurity and usable privacy and security.

\section{Background and Related Work} \label{sec:related}
In this section, we summarize related work on drawing in the learning context, collaborative drawing, and drawing in the CS and cybersecurity domains. We then provide background on our course and project, in which we incorporated the hand drawing exercise that we surveyed our students about.

\subsection{Categories of Drawing in the Learning Context} \label{sec:drawing_types}
Quillin and Thomas suggested that drawing in learning can be representational or abstract~\cite{quillin15}. Representational drawing depicts objects and images, while abstract drawing depicts processes and relationships~\cite{quillin15}. They further classified drawing exercises into formative or summative. Formative exercises allow students to construct their own knowledge and enable targeted instructor feedback, while summative exercises allow students to communicate their knowledge and skills and enable instructors to evaluate students' performance~\cite{quillin15}. Drawings can be made using various media and tools~\cite{dechamps25}. They can be made on traditional paper with pencils and markers, or on modern digital devices such as tablets~\cite{quillin15}. Moreover, drawing can be an individual or collaborative exercise where multiple students contribute to the same drawing~\cite{park21,tytler20}. Drawings can vary in the extent to which they are generated by the learner~\cite{quillin15}. This is especially true with the introduction of generative artificial intelligence (GenAI) tools such as ChatGPT~\cite{chatgpt26} and Gemini~\cite{gemini26}. Finally, drawing can be supported with different amounts of text~\cite{quillin15,hartel18}.

\subsection{Drawing to Learn} \label{sec:drawing_learn}
The drawing process is not merely an external representation of a static internal model~\cite{quillin15}. Studies suggest that drawing involves interactions between internal and external models. It involves selecting relevant information, structuring it, and finally integrating all into a mental model~\cite{mayer09,quillin15}. Drawing an external physical model can occur during or after creating the mental model~\cite{meter05,quillin15}. Drawing involves \quotes{observing, analyzing, reflecting, understanding and thinking}~\cite{adams17}. 

The cognitive process of drawing has the potential to enhance student learning~\cite{meter05}. Research suggests that drawing can foster various pedagogical goals. Ainsworth proposed five reasons why drawing should be considered an essential element in science education alongside writing, reading, and speaking: \begin{enumerate*} \item to reason in science, \item to improve engagement, \item as a learning strategy, \item to learn to represent in science, and \item to communicate~\cite{ainsworth11}\end{enumerate*}. Adams suggested four purposes that drawing can serve in learning: \begin{enumerate*} \item perception, \item communication, \item invention, and \item action~\cite{adams17}\end{enumerate*}. Drawing can reveal misconceptions~\cite{dikmenli10}, and help students understand complex concepts and systems~\cite{fish07}. Quillin and Thomas suggested that formative representational drawing exercises foster: \begin{enumerate*} \item active learning, \item observational skills, \item memorization, \item understanding of spatial relationships, and \item motivation/enjoyment of learning~\cite{quillin15}\end{enumerate*}.

Several studies employed drawing to understand people's perceptions about certain topics. Bowden et al. used drawing to understand people's perspectives about energy by responding to the question: \quotes{what does energy look like?}~\cite{bowden15}. Guillemin used drawing to understand illness conditions~\cite{guillemin04}. Hartel et al. used the \quotes{draw-and-write technique} to understand people's perception of information~\cite{hartel18}. Weber et al. asked university students enrolled in an elementary education program to draw a \enquote{teacher}, write about their drawing, and share and discuss their perception with their peers~\cite{weber95}. 

\subsection{Collaborative Drawing} \label{sec:collab_drawing}
Several studies suggested that collaborative drawing offers numerous benefits. Bracey et al.'s study of community college students studying cosmology suggested that collaborative drawing fosters an interactive learning environment in which students engaged with cosmology concepts~\cite{bracey17}. Tyler et al.'s study on year-7 students in a science class suggested that drawing in groups enables reasoning through and from drawing~\cite{tytler20}. Schwartz's experiments on secondary-school students while solving novel scientific problems in science suggested that working in groups encourages learners to draw more than individuals, explaining that with the knowledge exchange that occurs within groups~\cite{schwartz95}. Park et al. described the interaction process in collaborative drawing with the Plan\allowbreak‐Draw\allowbreak‐Evaluate (PDE) pattern, which facilitates the development of students' ideas through suggesting, visualizing, elaborating, agreeing, and disagreeing.~\cite{park21}.

\subsection{Drawing in the CS and Cybersecurity Domains} \label{sec:drawing_ups}
Several studies have used drawing in the Computer Science (CS) domain. Hansen et al. developed the Draw-A-Computer-Scientist-Test (DACST) to understand primary school students' conceptions and misconceptions about CS professionals~\cite {hansen17}. 
Moskal et al. used drawings to understand first-year college students' perceptions about programming. The students' drawings enabled them to identify recurrent patterns of artifacts, actors, activities, aspirations, and affect~\cite{moskal17}. Xu et al. used the \quotes{draw-a-picture} technique to analyze college students' perception of learning computer science~\cite{xu21}.

In the cybersecurity domain, the literature shows that multimedia (using more than one form of communication, which may include text and images~\cite{zhang21}), such as comics for example, has been used in cybersecurity awareness and education tools, especially for novice users (including students)~\cite{zhang21}. Drawing has also been used in research to gain insights into participants' knowledge, perceptions, and mental models of certain issues~\cite{distler21}. Kang et al. used mental models to examine people's knowledge about the Internet and how it affects their response to security and privacy risks~\cite{kang15}. Oates et al. employed illustrations in response to the question \quotes{what does privacy mean to you?} to identify mental models and metaphors of privacy to improve privacy tools, design, and communication for laypersons~\cite{oates18}. Sturdee et al. asked participants to draw cybersecurity concepts (risk, privacy, and trust)~\cite{sturdee21}. The generated drawings can improve cybersecurity visual vocabulary and communication~\cite{sturdee21}. 

To the best of our knowledge, we are not aware of any research that reports students' perspectives and gains from an in-class hand drawing exercise to conceptualize usable privacy and security research problems. This work aims to bridge this knowledge gap.

\subsection{Our Information and Computer Security Course and Project} \label{sec:course}
Our hand drawing exercise was conducted as an in-class activity incorporated in the Information and Computer Security course for undergraduate (BSc) students in the Information Systems Department at the Faculty of Computing and Information Technology, King Abdulaziz University, Saudi Arabia~\cite{kau26}, taught by the first author of this paper in 2025 and 2026. The course included a semester-long educational research project that accounts for 20\% of the student's final grade (out of 100), with an emphasis on usable privacy and security issues~\cite{distler21}. Project topics included deceptive patterns~\cite{driver26}, misinformation~\cite{altuncu26}, cyberbullying~\cite{campbell21}, smart homes and wearable device privacy~\cite{zallio19,madanian21,fienberg25}, and other related topics, which are updated over time. 

At the beginning of the course, students form project groups, ideally 4 to 5 students per group (with exceptions as needed), and select a research problem in usable privacy and security from a list of topics provided by the instructor. Each topic in the list is supplemented with a brief description and a reading list. Students can propose their own topic upon the instructor's approval, but this is the exception, not the norm. Each group is required to submit two preferred topics along with a specific initial research question and a motivation statement for each. The instructor revises the research questions and approves or requests revisions as needed. The instructor assigns topics and research questions so that each group has a unique research question for their project. 

The project spans a full semester. 
The project deliverables are a research report in the form of a short research paper (6 pages of main body) and a short presentation for classmates and the instructor. We omit the project's implementation fine-grained details as they are outside this paper's scope.

\subsection{Our Hand Drawing Exercise} \label{sec:drawing_ex}
Our hand drawing exercise took place during the conceptualization phase of the course's student research project, after students had selected a research question to investigate. To familiarize students with the drawing exercise, the instructor provides a few short, informal tutorials and examples showing how to draw a poster by hand~\cite{youtube_1,youtube_2,youtube_3,youtube_4}. In addition, starting from the second cohort, the instructor presents exemplary posters produced by previous students.

On the day of the drawing exercise, the instructor provides each group with a flip chart, colored markers, pencils, erasers, and the necessary tools for hand drawing. Students collaboratively draw in groups (each group comprising the research project members) and are given approximately 30 to 50 minutes to complete their drawing (though this timing varied by class across cohorts). Groups requesting extended time after class upon request are normally allowed to have more time. However, all groups have to submit their drawing during the class day.~\Cref{sec:stu_posters} shows samples of students' generated drawings. After completing the drawing exercise, each group briefly presents and explains their poster to their classmates and instructor. 

The first author was introduced to this exercise in the Usable Privacy and Security course at Carnegie Mellon University (CMU) taught by Lorrie Faith Cranor in 2023.

\section{Methods}
\subsection{Study Scope}
The hand drawing exercise was a learning activity conducted as part of the student research project in the Information and Computer Security course. It took place before we designed the survey of students’ perspectives. We did not implement the drawing exercise as a research intervention for this paper. We do not use the drawings as research data for this paper. Instead, this paper examines students’ perspectives on the exercise and reports our experience implementing it as a learning activity. In what follows, we describe the research methodology used to collect and analyze students’ perspectives. Background on our course, project, and the hand drawing exercise is provided in~\Cref{sec:course} and~\Cref{sec:drawing_ex}. 

\subsection{Recruitment} 
We invited the 113 students who completed or were currently enrolled in the Information and Computer Security course taught by the first author of this paper in 2025 and 2026. We screened for students who completed or were currently enrolled in the Information and Computer Security course taught by the first author of this paper starting from 2024\footnote{The first Information and Computer Security course taught by the first author that included the hand drawing exercise was in Fall 2025 of the 2024/2025 academic year, hence we used \enquote{starting from 2024} in the survey screening question.}, participated in the collaborative hand drawing exercise, and were able to remember the experience well enough to answer questions about the exercise. The survey was distributed in 2026. 
There was no compensation for participation. However, participants were informed in the invitation email and consent form that they can optionally enter a ballot for a 100 Saudi Riyals ($\approx$ 27 US Dollars) Amazon online store voucher, which will be given to 15 randomly selected participants who complete the survey. 

\subsection{Survey}
To gain insights into students' perspectives on the hand drawing exercise, we created a survey entitled \quotes{Students’ Perspectives on the hand drawing exercise in the Information and Computer Security Course Project (CPIS-312)}. The survey aimed to gain insights into the extent to which the drawing exercise was helpful or unhelpful as a pedagogical tool to support students in their usable privacy and security research project. 

The survey included both multiple-choice and open-ended short-answer questions. The survey was designed using Google Forms in Arabic, the native language of the participants and researchers.  

The survey begins with a consent form that includes the study title, purpose, procedures, risks, and benefits. It then asks eligibility questions, presents an example of the conceptualization poster to remind participants of the exercise, and asks whether they recall the experience sufficiently to answer questions about their perspective on it. The survey quits if the participant does not meet the eligibility criteria or cannot recall the experience well enough. Otherwise, the survey proceeds to the next section. 

After a question about the participant's age, the survey asks about the hand drawing experience, comprising eight 5-point Likert-scale questions, yes/no questions, and open-ended questions. See \autorefappendix{app:survey} for the survey. The survey first asks how helpful or unhelpful the exercise was with respect to five aspects: \begin{enumerate*} \item developing the research question, \item clarifying the student's conception of the project's idea, \item motivating the student to work on the project, \item developing a positive relationship with the team members, \item fostering the student's sense of ownership over the project\end{enumerate*}. It then asks questions about the exercise's joyfulness and the extent to which they would recommend it to other students taking the same course. It then asks participants a few questions about their willingness to have such an exercise in other courses at the Faculty of Computing and Information Technology, what these courses are, and how the drawing exercise can be useful in them. The survey then asks whether participants experienced any downsides during the drawing exercises and, if so, what they were. Finally, it asks participants for any further comments before closing the survey.

After submitting the survey, participants were provided with a link to enroll in the Amazon voucher ballot if they wished. The link takes participants to a separate survey disconnected from the study survey, which requests their contact details while maintaining participants' anonymity in the study's survey. 

\subsection{Ethics}
We obtained ethical approval from our institution's (King Abdulaziz University) ethics committee to conduct our survey study. All participants were aged 18 years or older and provided consent before taking part in the survey. Participants were informed that their participation in the survey is voluntary and that they can withdraw at any time before submitting the survey by closing the web browser tab. We did not collect any personally identifying data, such as names or email addresses. The voucher survey was separate from the study's survey, so we cannot de-identify participants who wanted to enroll in Amazon's voucher ballot. 

\subsection{Data Analysis}
Our survey contains both quantitative and qualitative data. Quantitative data were analyzed using descriptive statistics~\cite{kaur18}. Qualitative analysis was performed by two members of the research team using template analysis, a form of thematic analysis that involves both inductive and deductive coding~\cite{king24,brooks14}. One researcher acted as a primary coder and the second as a reviewer. The primary coder read all the open-ended answers, created a codebook, and coded the data. The reviewer reviewed all codes, and any disagreements were discussed jointly until both researchers reached an agreement on all codes.

\subsection{Limitations}
Our study has several limitations. First, due to the gender-segregation policy at our university, our sample is limited to female students. Moreover, our sample is from Saudi Arabia, which may be culturally, economically, and politically different from the WEIRD (Western, Educated, Industrialized, Rich, and Democratic) samples~\cite{ayako24}. This may limit the generalizability of our results. Nevertheless, it does not diminish the value of sharing our experience and findings. Second, our participants are prone to social desirability, where students may provide more positive answers, believing that this is the desirable response for the university and instructor. To mitigate this bias, the survey did not collect any identifiable data such as name, email, cohort, or course section, or whether the student has completed or is currently enrolled in the course. Fourth, some participants may be prone to recall bias, as our survey was distributed to former students, including those who completed the exercise two semesters earlier. To mitigate this bias, we included a question at the beginning of the survey that reminded students about the hand drawing exercises, accompanied by a description and a photo of a poster from one of the groups. We asked the students if they recalled the experience well enough to answer questions about their perspective on the exercise before proceeding to the survey questions. Participants who answered \quotes{no} did not proceed to the rest of the survey, and no questions were shown to them. Finally, similar to most voluntary surveys, our survey is prone to self-selection bias. That is, students who chose to participate may differ from those who did not, and thus may not be representative of all students who participated in the drawing exercise.

\begin{table}[h!]
\caption{Participants' sentiments on the drawing exercise.}
\label{tab:results}
\centering
\footnotesize
\begin{tabular}{p{5cm}|ll}
    \toprule
    \multicolumn{3}{c}{\textit{N = 64}} \\
    \midrule
    \textbf{Q\#7} Developing the research question & \# & \% \\
    \hline
        \quad Very helpful & 38 & (59\%)\\
        \quad Helpful & 19 & (30\%)\\
        \quad Neutral & 5 & (8\%)\\
        \quad Not helpful & 1 & (2\%)\\
        \quad Not at all helpful & 1 & (2\%)\\
    \hline
    \textbf{Q\#8} Clarifying the student's conception of the project idea & \# & \% \\
    \hline
        \quad Very helpful & 48 & (75\%)\\
        \quad Helpful & 12 & (19\%)\\
        \quad Neutral & 3 & (5\%)\\
        \quad Not helpful & 0 & (0\%)\\
        \quad Not at all helpful & 1 & (2\%)\\
    \hline
    \textbf{Q\#9} Increasing motivation to work on the project & \# & \% \\
    \hline
        \quad Very helpful & 47 & (73\%)\\
        \quad Helpful & 14 & (22\%)\\
        \quad Neutral & 2 & (3\%)\\
        \quad Not helpful & 1 & (2\%)\\
        \quad Not at all helpful & 0 & (0\%)\\
    \hline
    \textbf{Q\#10} Developing positive team relationship & \# & \% \\
    \hline
        \quad Very helpful & 52 & (81\%)\\
        \quad Helpful & 10 & (16\%)\\
        \quad Neutral & 2 & (3\%)\\
        \quad Not helpful & 0 & (0\%)\\
        \quad Not at all helpful & 0 & (0\%)\\
    \hline
    \textbf{Q\#11} Fostering a sense of project ownership & \# & \% \\
    \hline
        \quad Very helpful & 33 & (52\%)\\
        \quad Helpful & 24 & (38\%)\\
        \quad Neutral & 6 & (9\%)\\
        \quad Not helpful & 0 & (0\%)\\
        \quad Not at all helpful & 1 & (2\%)\\
    \hline
    \textbf{Q\#12} The hand drawing exercise was enjoyable & \# & \% \\
    \hline
        \quad Very enjoyable & 55 & (86\%)\\
        \quad Enjoyable & 7 & (11\%)\\
        \quad Neutral & 2 & (3\%)\\
        \quad Not enjoyable & 0 & (0\%)\\
        \quad Not at all enjoyable & 0 & (0\%)\\
    \hline 
    \textbf{Q\#13} Recommend the hand drawing exercise to future students in the Information and Computer Security course & \# & \% \\
    \hline
        \quad Strongly recommend it & 51 & (80\%)\\
        \quad Recommend & 9 & (14\%)\\
        \quad Neutral & 3 & (5\%)\\
        \quad Do not recommend it & 0 & (0\%)\\
        \quad Do not recommend it at all & 1 & (2\%)\\
    \hline
    \textbf{Q\#14} Include hand drawing exercises in other courses & \# & \% \\
    \hline
        \quad Would very much like to & 37 & (58\%)\\
        \quad Would like to & 17 & (27\%)\\
        \quad Neutral & 5 & (8\%)\\
        \quad Would not like to & 3 & (5\%)\\
        \quad Would not like to at all & 2 & (3\%)\\
    \hline 
   \textbf{Q\#16} Downsides faced throughout the hand drawing exercise & \# & \% \\
    \hline
        \quad Yes & 3 & (5\%)\\
        \quad No & 61 & (95\%)\\
    \bottomrule
\end{tabular}
\end{table}

\section{Results}
\begin{figure*}[!t]
    \centering
    \includegraphics[width=0.85\textwidth]{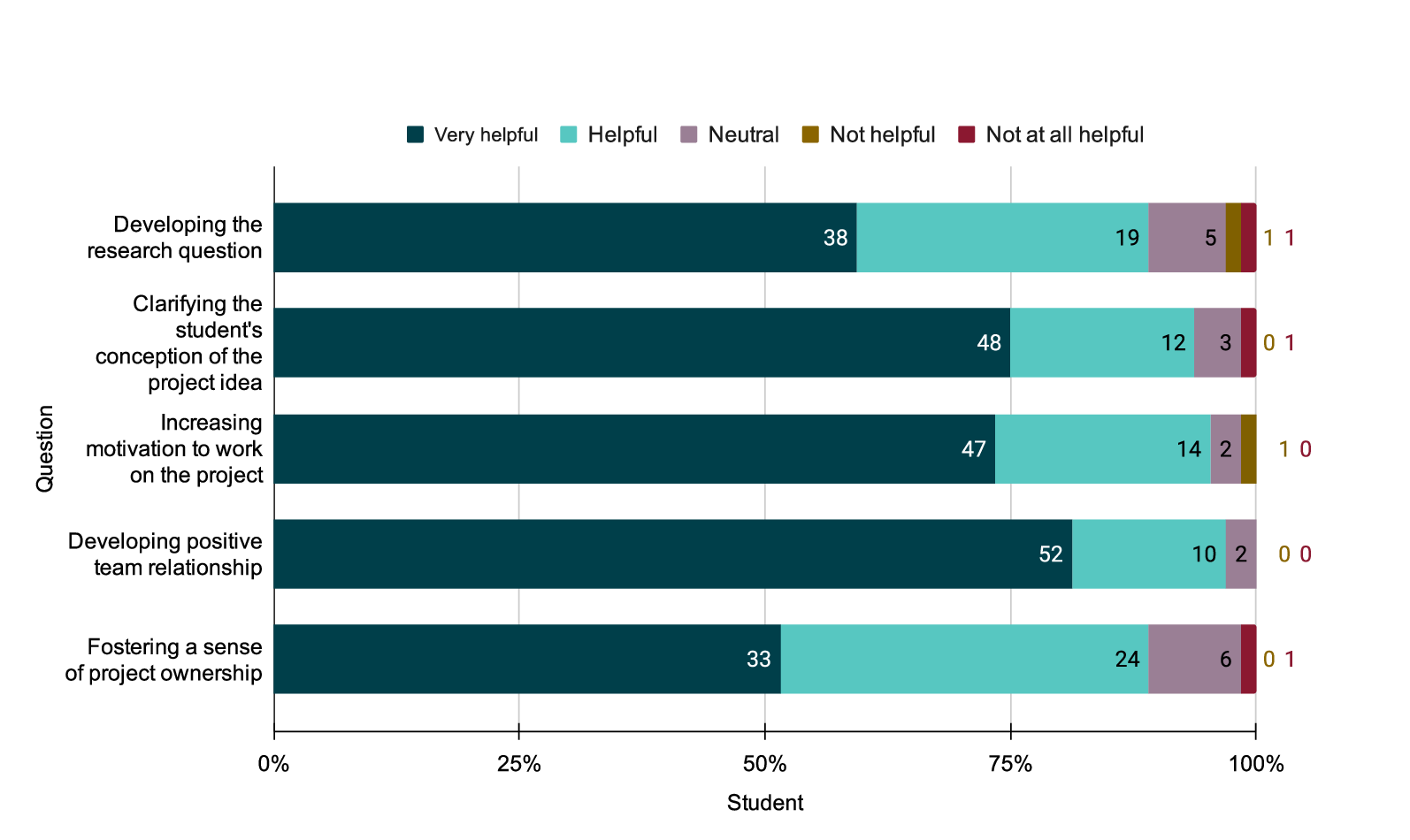}
    \vspace{-0.5cm}
    \caption{Students' perspectives regarding the helpfulness of the drawing exercise in their research project, ordered according to the question order in the survey.}
    \label{fig:results}
\end{figure*}
In this section, we summarize our results. In what follows, for summary and readability, we report the 5-point Likert-scale results by merging the two lowest and the two highest response options. For example, in the helpfulness questions, we merge both \enquote{very helpful} and \enquote{helpful} as \enquote{helpful}, and \enquote{not helpful} and \enquote{not at all helpful} as \enquote{not helpful}. The same goes for the rest of the 5-point Likert scales. Full-scale results are provided in~\Cref{tab:results}. We report frequencies as \#/\#, where the top number denotes the number of occurrences, and the bottom denotes the total number of participants considered, followed by the percentage (\#\%). Participants answered open-ended questions in Arabic, English, or both. We translated Arabic answers and reported all quotes in English. We denote participants as P\_\# and in qualitative themes, we report the number of participants who mentioned the theme in parentheses (\# students) to give insights into how frequent it is.   

\subsection{Participants}
Out of the 113 invited participants, 64 completed the survey. All participants were officially registered as female at our university, due to gender segregation. Thus, we did not ask about gender. In terms of age, 62/64 (97\%) were aged 18 to 24 years, and 2/64 (3\%) were aged 25 to 34 years.

\subsection{Student Sentiments Towards the Hand Drawing Exercise}
\subsubsection{Helpfulness}
As depicted in~\Cref{fig:results}, our results show that, overall, the hand drawing exercise was perceived by students as helpful in several aspects. It was most helpful in developing a positive relationship between team members, as reported by 62/64 (97\%), and in increasing students' motivation to work on the project, as reported by 61/64 (95\%). Regarding clarifying the student's conception of the project idea, 60/64 (94\%) reported it was helpful. The exercise was also perceived to be helpful, though to a lesser extent, in the development of the project's research question, as reported by 57/64 (89\%), and in fostering a sense of ownership of the project, as reported by 57/64 (89\%).

\subsubsection{Joyfulness}
The hand drawing exercise was experienced in a positive light, with 62/64 (97\%) of participants found it enjoyable, and none said it was not enjoyable. Additionally, the open-ended question about additional comments indicated joyfulness. That is, enjoyable was the most recurrent theme in the comments, mentioned by 16 participants, followed by interesting, mentioned by 7 participants. 

\subsubsection{Recommendedness}
Among our participants, 60/64 (94\%) said that they would recommend the hand drwing exercise to future students in the research project in the Information and Computer Security course. 

\subsubsection{Applicability to Other Courses}
When asked about their willingness to have such exercises in other courses at the Faculty of Computing and Information Technology, of the participants, 54/64 (84\%) said they would like to, 5/64 (8\%) would not like to, while 5/64 (8\%) were neutral. Following up with those who would like to, we asked them to name those courses and detail why and how drawing exercises can be helpful in these courses. The most notable courses by the students were: Principles of Human-Computer Interaction (15 students), Fundamentals of Data Networks (12 students), Software Engineering (9 students), Development of Information Systems Applications (4 students), Computer Architecture and Organization (2 students), and Professional Computing Issues (1 student). Moreover, 11 students said they would like to have hand drawing exercises in all courses.~\Cref{app:courses} lists all mentioned courses along with frequencies.

\subsubsection{Why Students Would Like to Have Hand Drawing Exercises in Other Courses}
Across multiple courses, the following reasons were repeatedly mentioned: it is interesting (8 students), helps in understanding the course's concepts (7 students) and content (4 students), improves creativity (5 students), and imagination (5 students). Some participants reiterated aspects we asked about in the survey, but now with respect to a range of courses, not only the Information and Computer Security course. Namely, they recommend it to other courses as it helps in conceptualizing project ideas (6 students), is enjoyable (7 students), fosters teamwork (2 students), a sense of ownership of the idea (1 student), and increases motivation (3 students).

\subsubsection{How Hand Drawing Exercises Can Be Helpful in Other Courses}
Some students provided details on how hand drawing can help in some courses. In the Fundamentals of Data Networks course, students mentioned that it will help them understand network concepts, including data flow, types of network topology, and the network layers model. In the Principles of Human-Computer Interaction course, students mentioned hand drawing in relation to interface design and scenario design. Students also mentioned that hand drawing can be useful in the Computer Architecture and Organization course for understanding the computer's internal structure and how commands are executed internally. In the Professional Computing Issues course, which contains a debate between student groups on controversial issues related to modern technologies, students suggested that hand drawing can strengthen the sense of connection to the cause they advocate and increase their motivation. In both Software Engineering and IS Applications Design and Development courses, students mentioned that hand drawing helps with imagination and understanding concepts.

\subsubsection{Downsides to the Hand Drawing Exercise} \label{sec:downside}
We asked our participants if they faced downsides throughout the hand drawing exercise. Only 3/64 (5\%) answered \quotes{yes}, while 61/64 (95\%) indicated that they did not face any. In a follow-up question, we asked those who faced drawbacks to elaborate. All 3 answers centered on doubts about their drawing skills. One participant added that she was concerned that this may affect the clarity of their concept.
\section{Summary of Results and Discussion}
\subsection{Student Gain from Collaborative Hand Drawing}
Our survey results show that students found the collaborative hand drawing exercise to conceptualize usable privacy and security research problems helpful at several levels: developing a positive relationship with their team members (97\%), increasing student motivation to work on the project (95\%), clarifying the student's conception of the project's idea (94\%), developing the research question (89\%), and fostering their sense of ownership of the project (89\%).

Our results support and complement previous work that suggests that (collaborative) hand drawing to learn supports learning, communication, and collaboration~\cite{meter05,ainsworth11,adams17,quillin15,bracey17,tytler20,schwartz95,park21}. Our study presents a case study in the field of cybersecurity education, and we are not aware of previous work that examined this.

\subsection{Negative Side of Drawing and Mitigation}
The only downside mentioned by our students, when asked whether they faced any, was doubts about their drawing skills. Previous work has also noted this particular concern among participants of drawing exercises~\cite{abegglen18,hartel18}. However, the percentage in our study is comparatively low. We suggest that introducing the exercise before the activity day and providing drawing educational materials might have helped reduce this concern among our participants. As we described earlier in~\Cref{sec:drawing_ex}, before the exercise day, we provided students with educational materials~\cite{youtube_1,youtube_2,youtube_3,youtube_4}  and starting with the second cohort, the instructor showed them exemplary posters created by former students to familiarize them with the task. Although we have not asked students about the effect of these materials, we anticipate that these steps helped prepare students to draw confidently and, consequently, contributed to the success of the exercise.

\subsection{Drawing with AI as a Collaborator}
We informed the students about the exercise before the exercise day and did not set out rules regarding the sources they could use, such as the Internet or AI tools such as ChatGPT~\cite{chatgpt26} or Gemini~\cite{gemini26}. After completing the study, we contacted the students and asked them whether their drawings were derived from or inspired by third-party images from the Internet or AI tools. We found a variety of approaches, ranging from completely original drawings created without outside sources, to drawings that reproduced third-party Internet images of common objects (e.g., computers, locks, hackers), to AI-assisted drawings generated from a user-provided concept and subsequently refined and drawn by hand. 

The use of AI as a collaborator was an unexpected yet interesting aspect that raises new open questions worth discussing, such as the effect of using AI on the quality of the resulting drawings, as well as on students' gains from the exercise. Since this study does not aim to analyze students' drawings, we leave this to future work.
\section{Conclusion}
In this paper, we shared our experience and presented students' perspectives on incorporating a hand drawing exercise to conceptualize usable privacy and security research problems. We surveyed 64 students who participated in a hand drawing exercise as part of students' research project on usable privacy and security in the Information and Computer Security course for undergraduate students at King Abdulaziz University. The survey responses suggest that the drawing exercise was perceived positively in all the examined aspects. It was perceived helpful at all examined aspects, enjoyable, and the students recommended the drawing exercise to other students in the same course and students would like to have such exercises in other computing courses.
\section{Acknowledgment}
We thank the survey participants for their time and insights. Eman Alashwali thanks Lorrie Faith Cranor for introducing her to the drawing exercise and the semester-long research project structure through the Usable Privacy and Security course at Carnegie Mellon University in 2023. The implementation of the hand drawing exercise at King Abdulaziz University that is reported in this paper, including the survey design, analysis, and results, is the authors' own. 

\begin{sloppypar}
\bibliographystyle{ACM-Reference-Format-1}
\bibliography{refs}
\end{sloppypar}
\appendix
\section{Samples of Student-Generated Drawings} \label{sec:stu_posters}
In what follows, we list selected student-generated hand drawings. We obtained retrospective written consent from the students to publish their drawings for illustration. Student names in the captions 
were added as students provided them in the posters. We complemented or corrected names as needed based on names submitted on the consent form. We redacted (using white color not to distort the drawing) students' IDs and section names from posters, if any, for privacy and security reasons. 
We disclose use of AI tools or third-party images in captions as reported by students in the consent form for publishing their drawings and private email communications. 
\begin{figure*}[t!]
    \centering
    \begin{minipage}[t]{0.4\textwidth}
        \centering
        \includegraphics[width=0.8\textwidth]{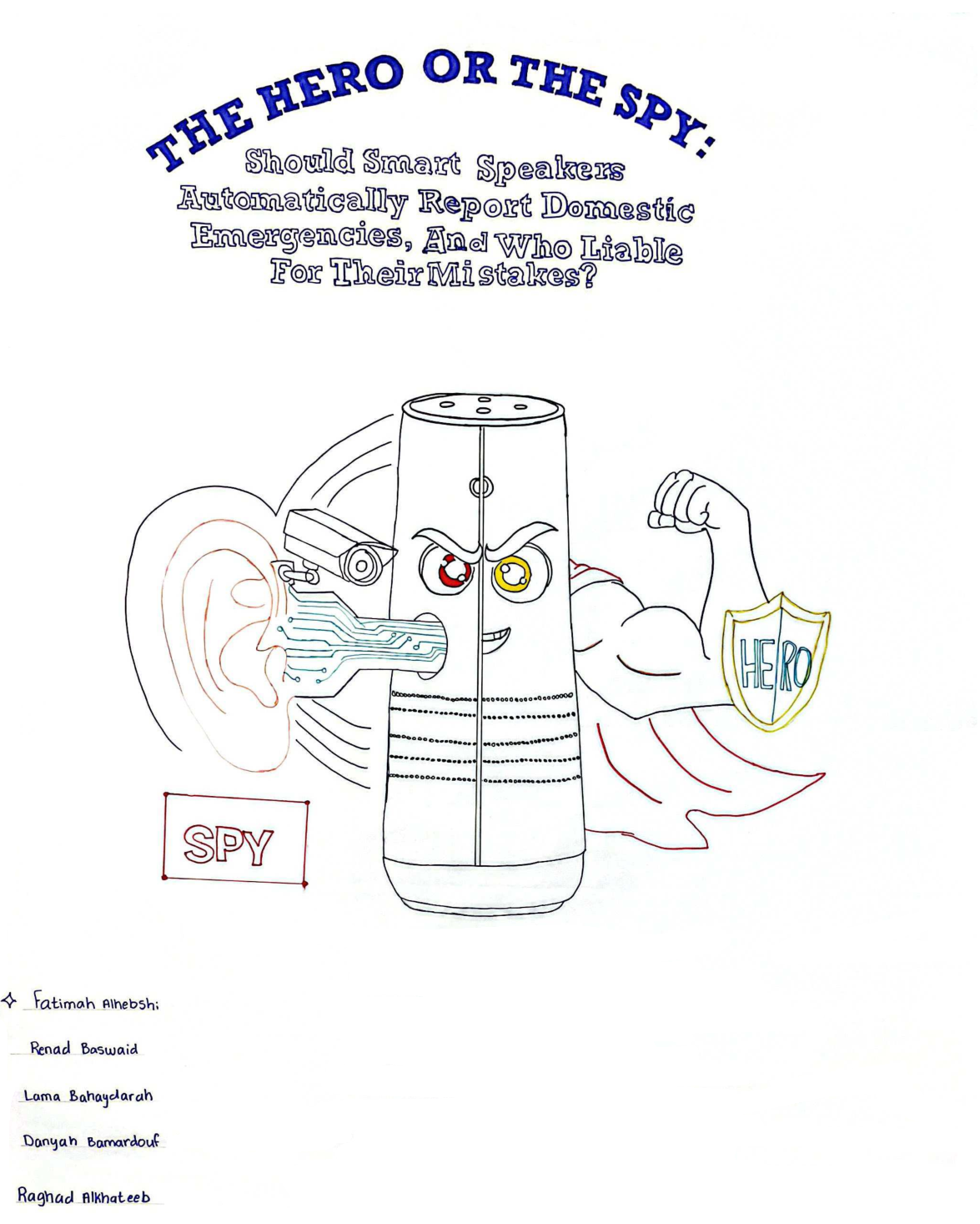}
        \caption{Drawing by: Fatimah Alhebshi, Renad Baswaied, Lama Bahaydarah, Danyah Bamardouf, and Raghad Alkhateeb. Text: \quotes{THE HERO OR THE SPY: Should Smart Speakers Automatically Report Domestic Emergencies, [a]nd Who [is] Liable For Their Mistakes?}, 2026. The students acknowledged using AI (Gemini) to assist in generating the drawing, but stated that the concept is their own.}
        \label{fig:subfig1}
    \end{minipage}
\hfill
    \begin{minipage}[t]{0.4\textwidth}
        \centering
        \includegraphics[width=0.8\textwidth]{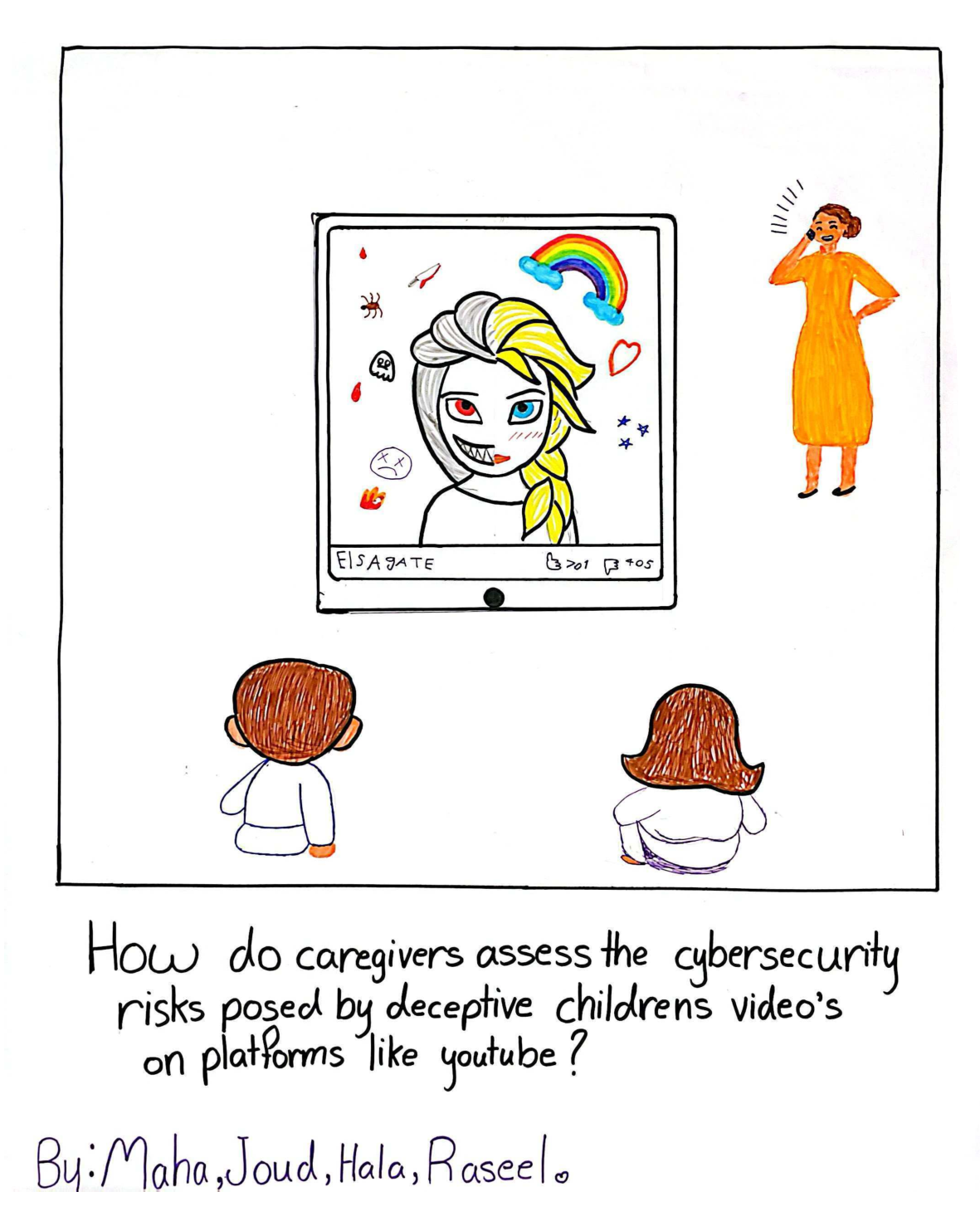}
        \caption{Drawing by: Maha Al-alsheikh, Joud Alzahrani, Hala Hakami, and Raseel Algonaisi. Text: \quotes{How do caregivers assess the cybersecurity risks posed by deceptive childrens video's on platforms like youtube?}, 2026. The students acknowledged reproducing images of common objects from third-party web sources (for which we do not have references) to assist in generating the drawing, but stated that the concept is their own.}
        \label{fig:subfig2}
    \end{minipage}
\end{figure*}
\begin{figure*}[t!]
    \begin{minipage}[t]{0.4\textwidth}
        \centering
        \includegraphics[width=0.8\textwidth,keepaspectratio]{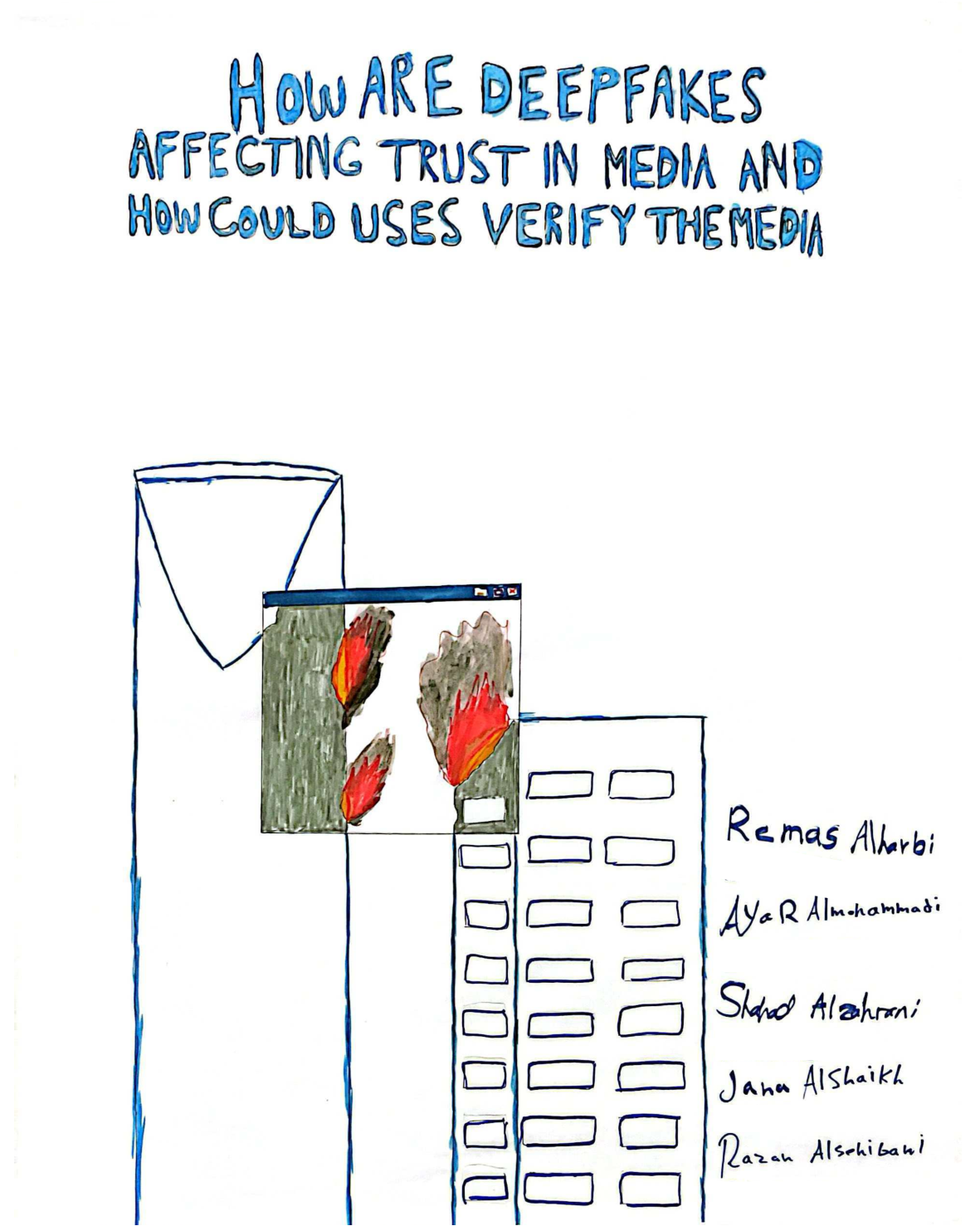}
        \caption{Drawing by: Remas Alharbi, Ayar Almohammadi, Shahad Alzahrani, Jana Alshaikh, and Razan Alsehibani. Text: \quotes{HOW ARE DEEPFAKES AFFECTING TRUST IN MEDIA AND HOW COULD USE[R]S VERIFY THE MEDIA} (we redacted student university IDs), 2026. The students acknowledged reproducing images of common objects from third-party web sources (for which we do not have references) to assist in generating the drawing, but stated that the concept is their own.}
        \label{fig:subfig3}
    \end{minipage}
\hfill
    \begin{minipage}[t]{0.4\textwidth}
        \centering
        \includegraphics[width=0.8\textwidth,keepaspectratio]{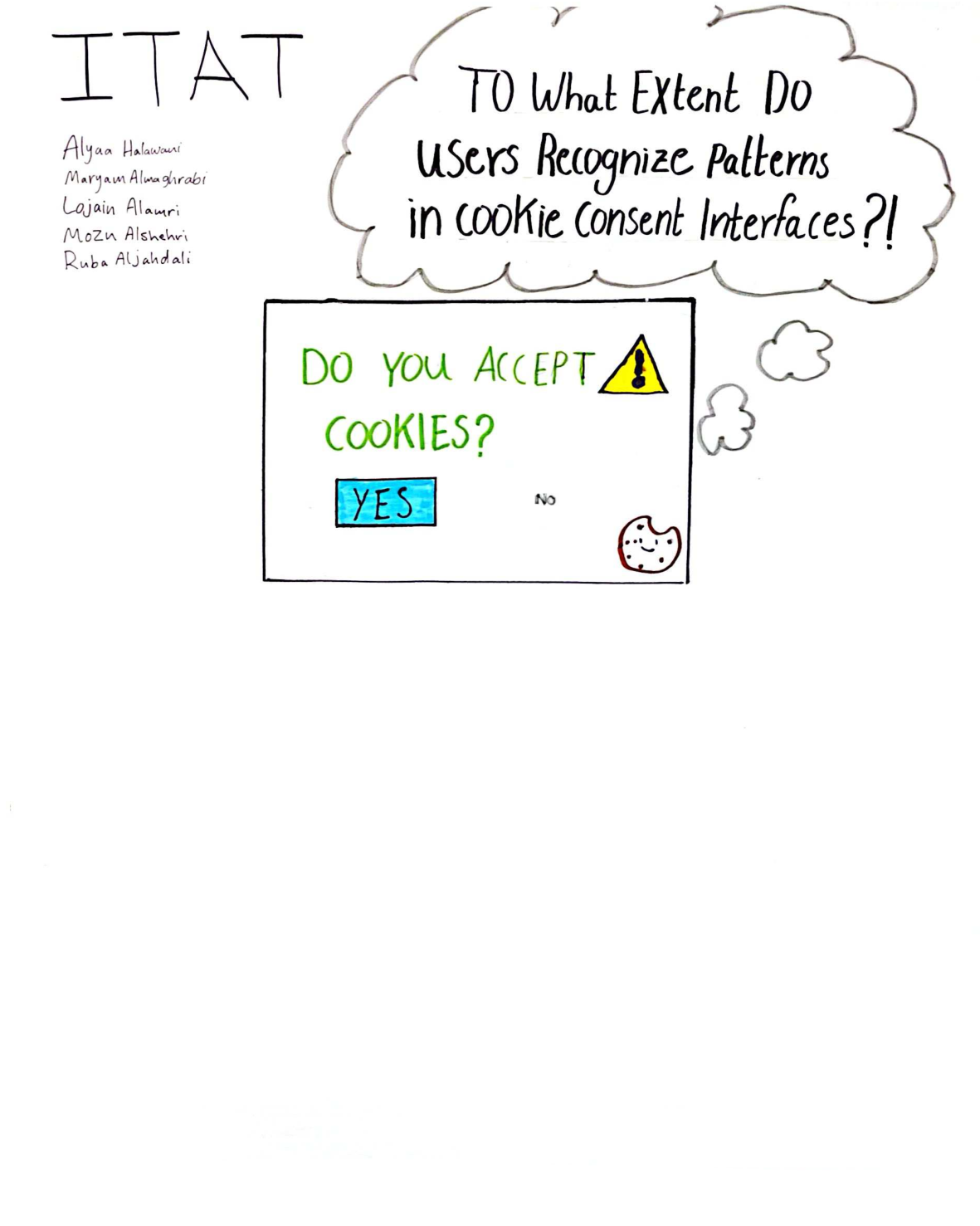}
        \caption{Drawing by: Alyaa Halawani, Maryam Almaghrabi, Lujain Alamri, Mozn Alshehri, and Ruba Aljahdali. Text: \quotes{To What Extent Do Users Recognize Patterns in [C]ookie Consent Interfaces?}, 2025. }
        \label{fig:subfig4}
    \end{minipage}
    \label{fig:four_subfigures}
\end{figure*}
\begin{figure*}[t!]
    \begin{minipage}[t]{0.42\textwidth}
        \centering
        \includegraphics[width=0.8\textwidth,keepaspectratio]{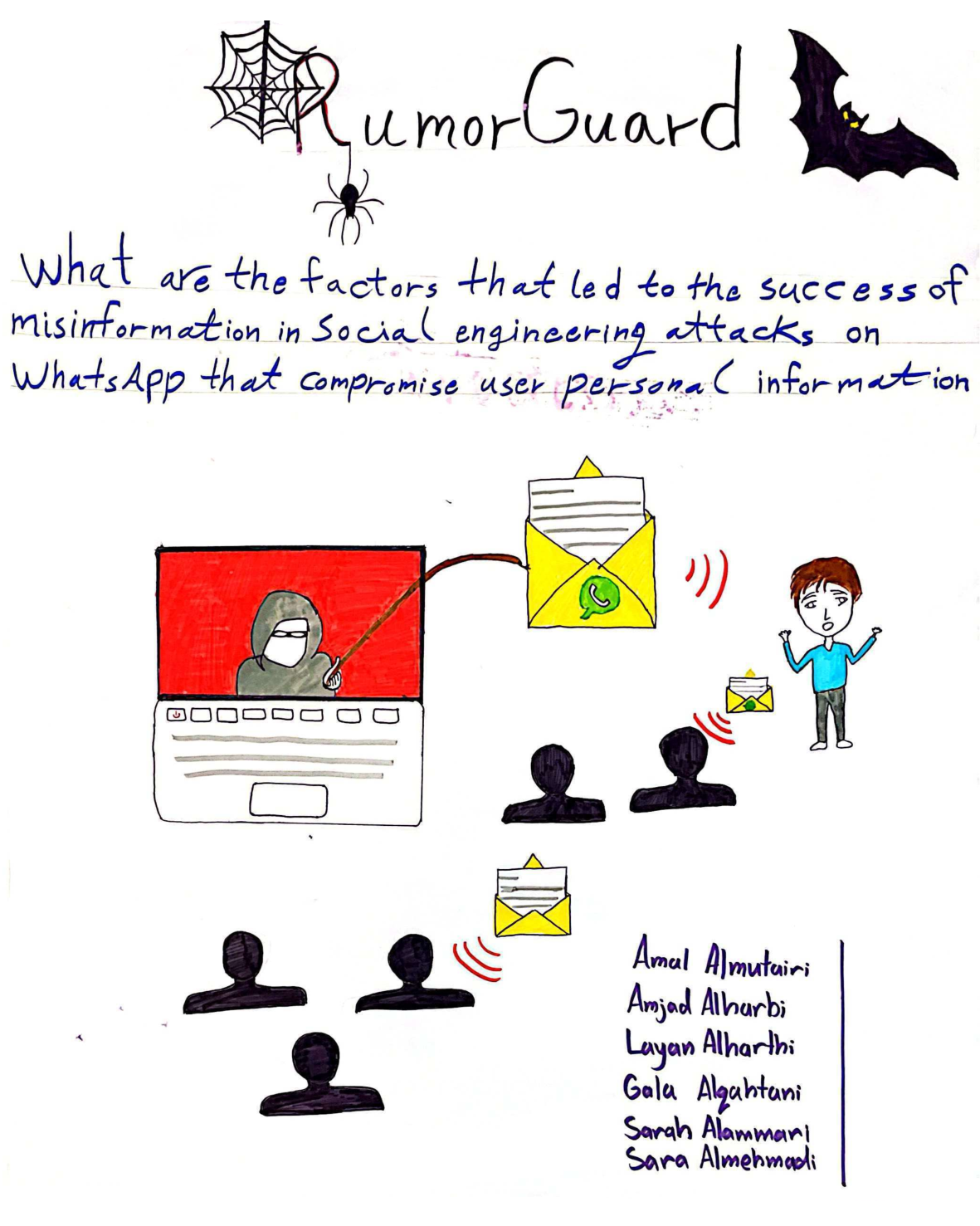}
        \caption{Drawing by: Amal Almutairi, Amjad Alharbi, Layan Alharthi, Gala Alqahtani, Sarah Alammari, and Sara Almehmadi. Text: \quotes{What are the factors that led to the success of misinformation in social engineering attacks on WhatsApp that compromise user personal information}, 2025. The students acknowledged reproducing images of common objects from third-party web sources (for which we do not have references) to assist in generating the drawing, but stated that the concept is their own.}
        \label{fig:subfig5}
    \end{minipage}
\hfill
    \begin{minipage}[t]{0.42\textwidth}
        \centering
        \includegraphics[width=0.8\textwidth,keepaspectratio]{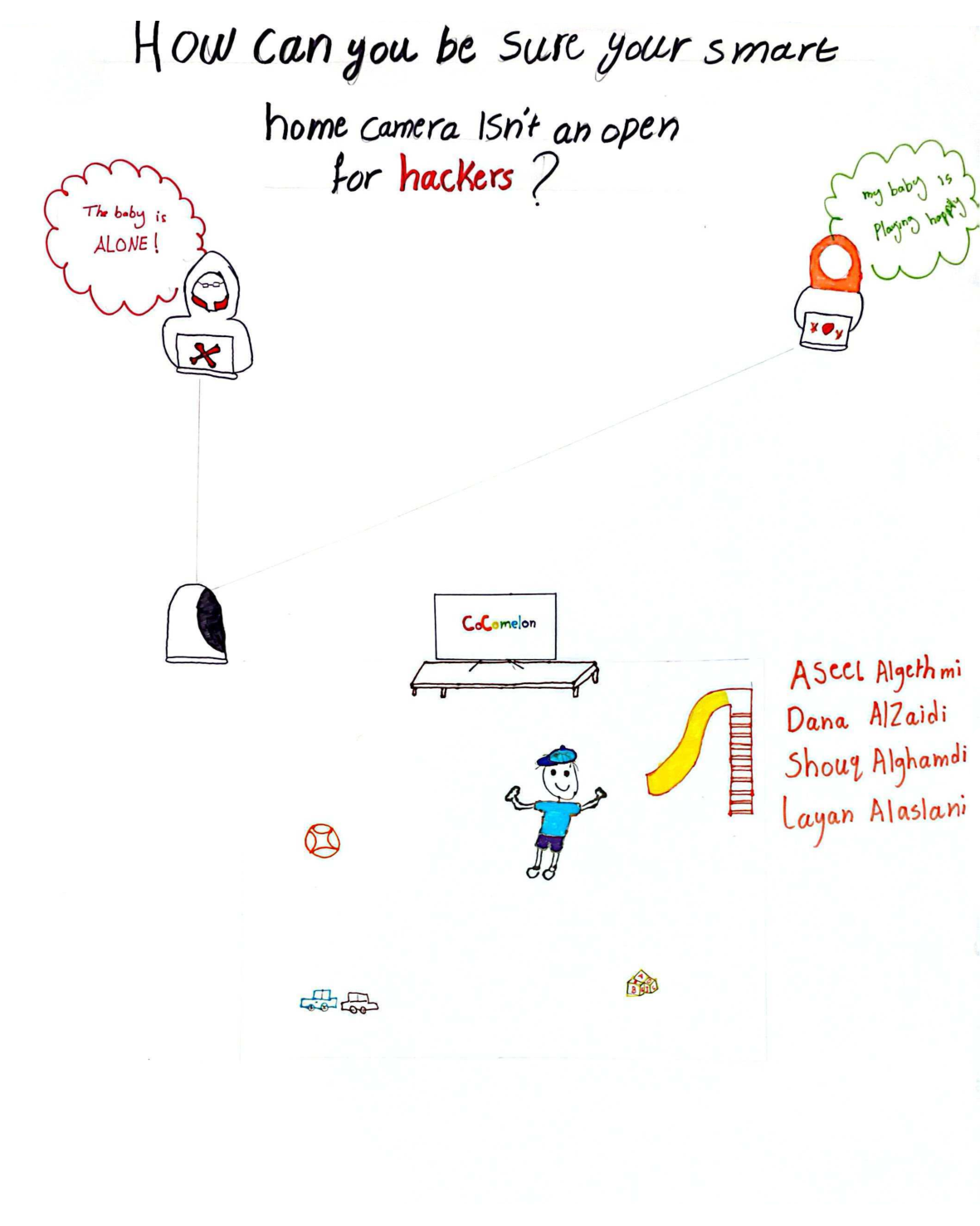}
        \caption{Drawing by: Aseel Algethami, Dana Alzaidi, Shouq Alghamdi, and Layan Alaslani. Text: \quotes{How can you be sure your smart home camera isn't [] open for hackers?}, 2025.}
        \label{fig:subfig6}
    \end{minipage}
    \label{fig:four_subfigures}
\end{figure*}
\begin{figure*}[t!]
    \begin{minipage}[t]{0.42\textwidth}
        \centering
        \includegraphics[width=0.8\textwidth,keepaspectratio]{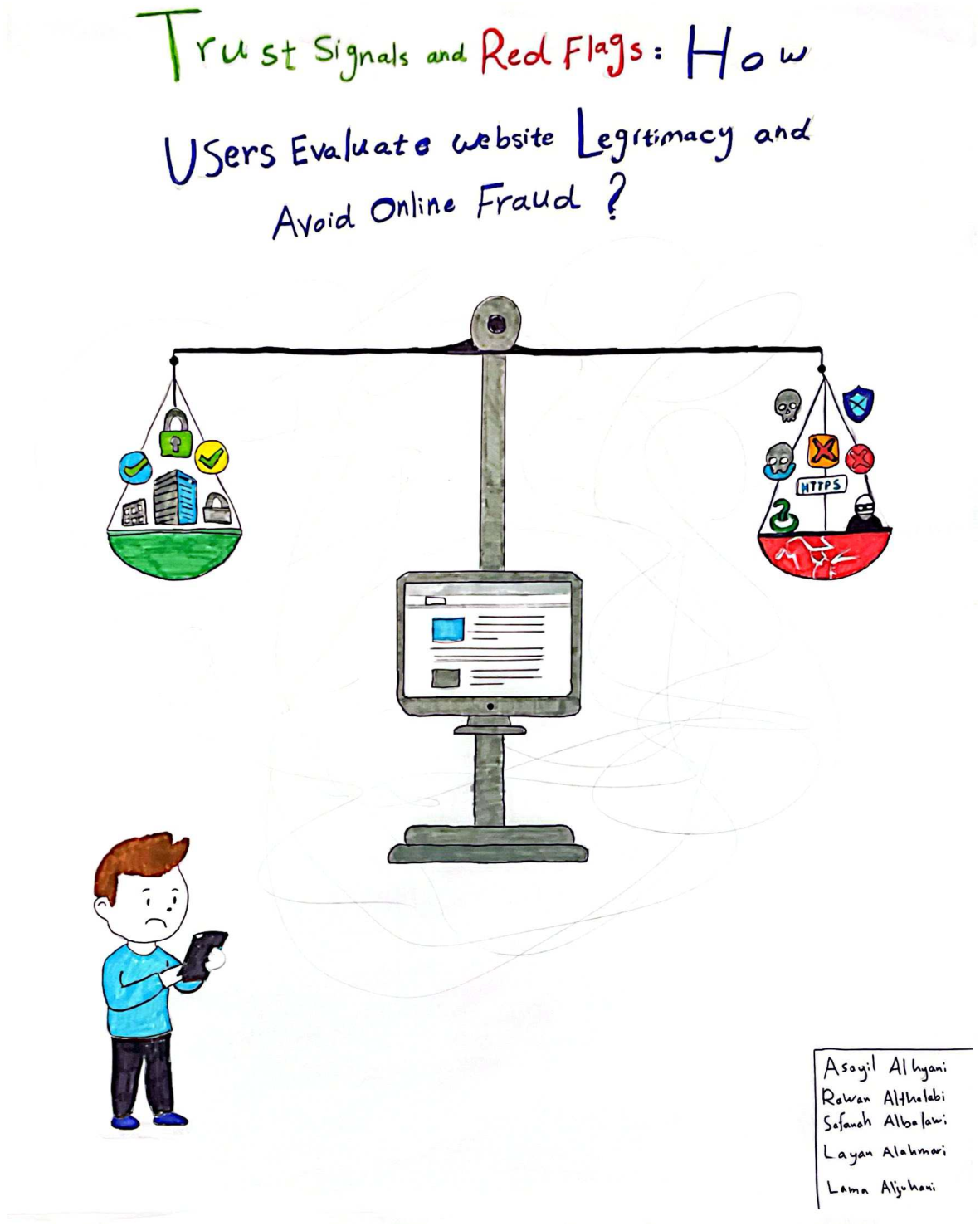}
        \caption{Drawing by: Asayil Allhyani, Rawan Althalabi, Safanah Albalawi, Layan Alahmari, and Lama Aljuhani. \quotes{Trust Signals and Red Flags: How Users Evaluate Website Legitimacy and Avoid Fraud?}, 2026. The students acknowledged using AI (Gemini/ChatGPT) to assist in generating the drawing, but stated that the concept is their own.}
        \label{fig:subfig8}
    \end{minipage}
\hfill
    \begin{minipage}[t]{0.42\textwidth}
        \centering
        \includegraphics[width=0.8\textwidth,keepaspectratio]{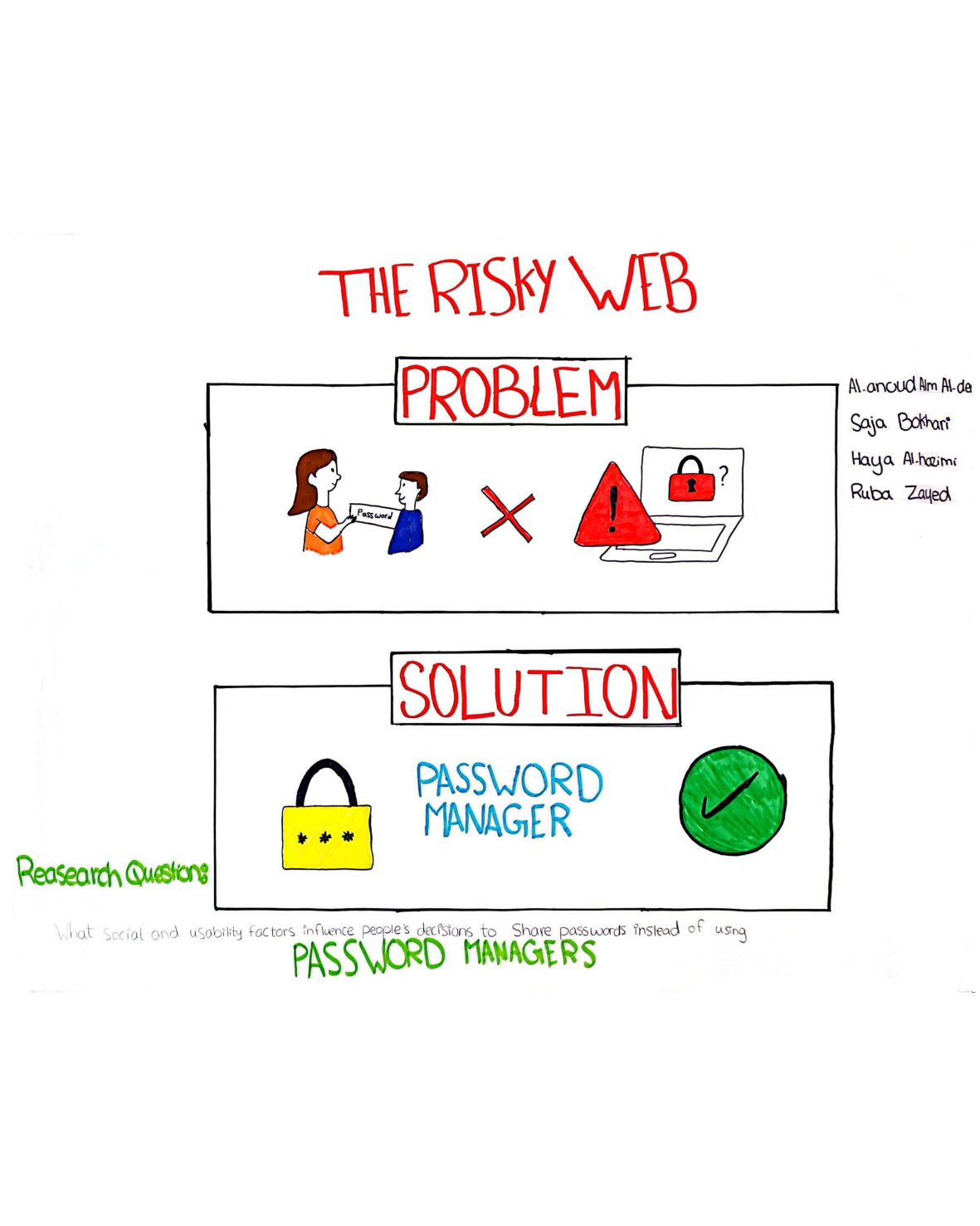}
        \caption{Drawing by: Alanoud Alm Aldeen, Saja Bokhari, Haya Alhazmi, and Ruba Zayed. Text: \quotes{What social and usability factors influence people's decisions to share passwords instead of using PASSWORDS MANAGERS}, 2025. The students acknowledged using AI (Gemini) to assist in generating the drawing, but stated that the concept is their own.}
        \label{fig:subfig8}
    \end{minipage}
\end{figure*}

\clearpage
\section{Main Survey} \label{app:survey}
In what follows, we provide the study survey (translated from Arabic to English). The original survey was in Arabic, as it is the native language of the participants and authors, enabling better understanding and expression. Text between square brackets \enquote{[ ]} was not part of the survey and is added by the authors for clarity. 

\noindent\textbf{[Survey title]}\\
\textbf{Students' Perspectives on the hand drawing Exercise in the Information and Computer Security Course Project (CPIS-312)} \newline

\noindent{\textbf{Introduction and Consent Form}}\newline
[Consent form here] \newline 

\noindent \textbf{Access code} \newline 
\noindent\textbf{Please enter the code number given to you by the research team to access the survey.} \\
\noindent [Open-text entry] \newline 

\noindent [If the access code is correct, the participant moves to the next section.] \\
\noindent \textbf{Consent} \newline 
\noindent\textbf{Q\#1: I would like to participate in this study and proceed to the survey}
\begin{itemize}
        \item I agree 
        \item I don't agree
    \end{itemize}
    
\noindent \textbf{Questions to Determine Eligibility For Participation}\\
\noindent\textbf{Q\#2: Are you 18 years old or older?}
    \begin{itemize}
        \item Yes 
        \item No
    \end{itemize}
    
\noindent\textbf{Q\#3: Did you take the Information and Computer Security course (CPIS-312) with Dr. Eman Alashwali starting 2024 or after?}
    \begin{itemize}
        \item Yes 
        \item No
    \end{itemize}
    
\noindent\textbf{Q\#4: Did you participate in-person in the hand drawing conceptualization poster exercise for the Information and Computer Security course project (CPIS-312)?}
    \begin{itemize}
        \item Yes 
        \item No
    \end{itemize}

\noindent [If Q\#1 to Q\#3 answers are \quotes{Yes}, the participant move to the next section.] \\
\noindent \textbf{Ability to Remember}\\
\noindent 
\textbf{Q\#5:} During the Information and Computer Security (CPIS-312) course with Dr. Eman Alashwali, in the early stages of the course project, you were asked to create a \textbf{conceptualization poster by hand} (\uline{see example below}) and present it to the students and professor in class. \newline
\noindent[A photo showing an example of a conceptualization poster was shown to the students here.]

\noindent \textbf{Do you remember this experience well enough to answer some questions about your perspective on this exercise?}
    \begin{itemize}
        \item Yes 
        \item No
    \end{itemize}

\noindent [If Q\#5 answer is \quotes{Yes}, the participant move to the next section.] \\
\noindent \textbf{Demographic Data}\\
\noindent \textbf{Q\#6: What is your age group (in years)?}
    \begin{itemize}
        \item 18 to 24 
        \item 25 to 34
        \item 35 to 44
        \item 45 to 54
        \item 55 to 64
        \item 65 or older 
    \end{itemize}

\noindent \textbf{Based on your personal experience with the hand drawing exercise to create the conceptualization poster in the Information and Computer Security (CPIS-312)} research project, answer the following questions (Click the \quotes{Next} button) \newline 

\noindent \textbf{Drawing Exercise Experience}\\
\textbf{Q\#7:} To what extent was the hand drawing exercise helpful or unhelpful to you \textbf{in developing the project's research question?}
    \begin{itemize}
        \item Very helpful
        \item Helpful
        \item Neutral
        \item Not helpful
        \item Not at all helpful
    \end{itemize}
    
\noindent \textbf{Q\#8:} To what extent was the hand drawing exercise helpful or unhelpful for you in \textbf{clarifying your conception of the project idea?}
    \begin{itemize}
        \item Very helpful
        \item Helpful
        \item Neutral
        \item Not helpful
        \item Not at all helpful
    \end{itemize}

\noindent \textbf{Q\#9:} To what extent did the hand drawing exercise help or unhelpful for you in \textbf{increasing your motivation to work on the project?}
    \begin{itemize}
        \item Very helpful
        \item Helpful
        \item Neutral
        \item Not helpful
        \item Not at all helpful
    \end{itemize}

\noindent \textbf{Q\#10:} To what extent was the hand drawing exercise helpful or unhelpful for you in \textbf{developing a positive relationship with your team members?}
    \begin{itemize}
        \item Very helpful
        \item Helpful
        \item Neutral
        \item Not helpful
        \item Not at all helpful
    \end{itemize}
    
\noindent \textbf{Q\#11:} To what extent was the hand drawing exercise helpful or unhelpful for you in \textbf{fostering your sense of ownership of the project?}
    \begin{itemize}
        \item Very helpful
        \item Helpful
        \item Neutral
        \item Not helpful
        \item Not at all helpful
    \end{itemize}

\noindent \textbf{Q\#12:} To what extent was the \textbf{hand drawing exercise enjoyable} to you?
    \begin{itemize}
        \item Very enjoyable
        \item Enjoyable
        \item Neutral
        \item Not enjoyable
        \item Not at all enjoyable
    \end{itemize}

\noindent \textbf{Q\#13:} To what extent \textbf{would you recommend or not recommend the hand drawing exercise} for students who will be working on a research project in the Information and Computer Security course (CPIS-312) in the future?
    \begin{itemize}
        \item Strongly recommend it
        \item Recommend it
        \item Neutral
        \item Do not recommend it
        \item Do not recommend it at all
    \end{itemize}
\hfill\\
\noindent\textbf{Incorporating hand drawing exercises into other courses} \newline 
\noindent \textbf{Q\#14:} To what extent would you like to \textbf{include hand drawing exercises in other courses} at the Faculty of Computers and Information Technology (FCIT)?
    \begin{itemize}    
    \item Would very much like to
    \item Would like to
    \item Neutral
    \item Would not like to
    \item Would not like to at all
    \end{itemize}

\noindent [If Q\#14 answer is \quotes{Yes}, participants move to the next section.] \\
\noindent \textbf{Other courses you would like hand drawing exercises to be included in} \newline 
\noindent \textbf{Q\#15:} You said that you would like to have hand drawing exercises included in other courses at the Faculty of Computers and Information Technology. Can you mention courses where such an exercise can be included? (write the course code followed by its number (e.g., CPIS-\#\#\#, CPCS- \#\#\#, CPIT- \#\#\#).

\noindent Please elaborate as much as possible, and explain \textbf{why you chose these courses} and \textbf{how these exercises could be helpful in these courses?} \newline 
\noindent [Open-text entry] \newline 

\noindent \textbf{Downsides}\\
\noindent\textbf{Q\#16:} Did you encounter any downsides \textbf{while working on the hand drawing exercise} to create the \textbf{conceptualization poster} for the project?
    \begin{itemize}
        \item Yes 
        \item No
    \end{itemize}

\noindent [If Q\#16 answer is \quotes{Yes}, participants move to the next section.] \\
\noindent \textbf{Listing the Downsides}\newline 
\noindent\textbf{Q\#17:} You said you encountered downsides \textbf{while working on the hand drawing exercise} to create the \textbf{conceptualization poster} for the project. \textbf{Please, elaborate and specify the downsides you encountered?}
\noindent [Open-text entry] \newline 

\noindent\noindent\textbf{Last word!}\newline \\
\noindent\textbf{Q\#18:} Is there anything you would like to tell us about regarding your experience in the \textbf{hand drawing exercise to create the conceptualization poster in the Information and Computer Security course (CPIS-312)?}\\
\noindent [Open-text entry] \newline 

\noindent\textbf{End of Survey.}\newline 
\noindent \textbf{Thank you!} \\

\noindent You have reached the end of the survey. Click the \quotes{Submit} button to submit your answers.

\noindent \textbf{\uline{If you would like to enter the drawing for one of 15 Amazon gift vouchers worth 100 SAR each,}} please copy the link that will appear on the next page and fill in your contact information.
Please note that the survey answers are separate from the contact information form. Your survey answers will remain anonymous.

\clearpage
\section{Extended Results} \label{app:courses}
\Cref{tab:courses} lists the courses our participants mentioned in their answers to \textbf{(Q\#15)}, along with their frequencies. 
\begin{table}[h!]
\caption{Courses students would like to have hand drawing exercise incorporated into with the frequencies each course was mentioned \textbf{(Q\#15)}.}
\label{tab:courses}
\centering
\small
\begin{tabular}{p{2cm}|ll}
\toprule
\multicolumn{3}{c}{\textit{N = 54}} \\
\hline 
\makecell[l]{Course code\\{[non-exclusive]}}  & Course name & \# \\
\hline 
CPIS-354 & Principles of Human-Computer Interaction & 15 \\
 CPIS-370 & Fundamentals of Data Networks            & 12 \\
All      & This code refers to all courses &	11 \\
 CPIS-250 & Software Engineering     	            & 9 \\
 CPIS-352 & IS Applications Design and Development &	4 \\
 CPIS-210 & Computer Architecture and Organization  &	 2 \\
 CPIS-220 & Principles of Information Systems &	2 \\
 CPIS-351 & IS Analysis and Architecture Design &	2 \\
 CPIS-357 & Software Quality and Testing &	2 \\
 CPIS-358 & Internet Applications and Web Programming & 	2 \\
 CPIS-360 & Advanced Information Systems Technologies &	2 \\
 CPIS-380 & Introduction to E-Business Systems &	2 \\
 CPIS-393 & Workplace Skills for Technology Specialist &	2 \\
 CPIS-240 & Database Management Systems (I) &	1 \\
 CPIS-428 & Professional Computing Issues &	1 \\
 CPIS-490 & Selected Topics in IS &	1 \\
 CPIS-222 & Principles of Operating Systems &	1 \\
\bottomrule
\end{tabular}
\end{table}

\end{document}